\documentclass[aip,rsi,reprint,amsmath,amssymb]{revtex4-1}

\usepackage{nicefrac}
\usepackage{graphicx}
\usepackage[shortlabels]{enumitem}
\usepackage{placeins}
\usepackage{tabularx}
\usepackage{hyperref}
\hypersetup{
    hidelinks,
    pdftitle={Correction of the VIR-Visible data set from the Dawn mission}}

\begin{document}


\title{Correction of the VIR-Visible data set from the Dawn mission} 



\author{B. Rousseau}
\email[]{batiste.rousseau@inaf.it}
\affiliation{IAPS-INAF, via Fosso del Cavaliere 100, 00133, Rome, Italy}
\author{A. Raponi}
\affiliation{IAPS-INAF, via Fosso del Cavaliere 100, 00133, Rome, Italy}
\author{M. Ciarniello}
\affiliation{IAPS-INAF, via Fosso del Cavaliere 100, 00133, Rome, Italy}
\author{E. Ammannito}
\affiliation{Italian Space Agency (ASI), Via del Politecnico, 00133, Rome, Italy}
\author{F. G. Carrozzo}
\affiliation{IAPS-INAF, via Fosso del Cavaliere 100, 00133, Rome, Italy}
\author{M. C. De Sanctis}
\affiliation{IAPS-INAF, via Fosso del Cavaliere 100, 00133, Rome, Italy}
\author{S. Fonte}
\affiliation{IAPS-INAF, via Fosso del Cavaliere 100, 00133, Rome, Italy}
\author{A. Frigeri}
\affiliation{IAPS-INAF, via Fosso del Cavaliere 100, 00133, Rome, Italy}
\author{F. Tosi}
\affiliation{IAPS-INAF, via Fosso del Cavaliere 100, 00133, Rome, Italy}



\date{\today}


\begin{abstract}
\vspace{0.2cm}
The following article has been accepted by Review of Scientific Instruments on 24 November 2019. After it is published, it will be found at \href{https://publishing.aip.org/resources/librarians/products/journals/}{this link}. \href{https://doi.org/10.1063/1.5123362}{doi: 10.1063/1.5123362}\vspace{0.5cm}\par
Data acquired at Ceres by the visible channel of the Visible and InfraRed mapping spectrometer (VIR) on board the NASA Dawn spacecraft are affected by the temperatures of both the visible (VIS) and the infrared (IR) sensors, which are respectively a CCD and a HgCdTe array. The variations of the visible channel temperatures measured during the sessions of acquisitions are correlated with variations in the spectral slope and shape for all the mission phases. The infrared channel (IR) temperature is more stable during the acquisitions, nonetheless it is characterized by a bi-modal distribution whether the cryocooler (and therefore the IR channel) is used or not during the visible channel operations. When the infrared channel temperature is high (175K, i.e. not in use and with crycooler off), an additional negative slope and a distortion are observed in the spectra of the visible channel. We developed an empirical correction based on a reference spectrum for the whole data set; it is designed to correct the two issues related to the sensor temperatures that we have identified. The reference spectrum is calculated to be representative of the global Ceres' surface. It is also made of data acquired when the visible and infrared channel temperatures are equal to the ones measured during an observation of the Arcturus star by VIR, which is consistent with several ground-based observations. The developed correction allows reliable analysis and mapping to be performed by minimizing the artifacts induced by fluctuations of the VIS temperature. Thanks to this correction, a direct comparison between different mission phases during which VIR experienced different visible and infrared channel temperatures is now possible.
\end{abstract}


\maketitle 


\section{Introduction \label{Intro}}

The targets of the NASA Dawn mission were the asteroid Vesta and the dwarf planet Ceres \cite{2007_Russell}. In September 2012, after one year of observations, the spacecraft left Vesta and started the exploration of Ceres in 2015, ending at the end of October 2018. The scientific payload of Dawn was made of three instruments designed to characterize the surface: the Visible and InfraRed mapping spectrometer \cite{2011_De_Sanctis} (VIR), the Framing Camera \cite{2011_Sierks} (FC) and the Gamma Ray and Neutron Detector \cite{2011_Prettyman} (GRaND).\par
This study is focused on the correction of the data acquired by VIR, which is composed of two spectral channels: the visible channel (VIS), which is a two-dimensional Charged Coupled Device (CCD) covering the wavelength range 0.25--1.07 $\mu$m and the infrared channel (IR), which is a two-dimensional HgCdTe array covering the range 1.02--5.09 $\mu$m with a nominal spectral sampling of 1.8 nm/band (432 bands) and 9.8 nm/band (432 bands) respectively. VIR channels may work together or independently (mainly according to the scientific planning), sharing the same optical path while using two different sensors close to each other\cite{2011_De_Sanctis}. However, while the CCD sensor of the VIS channel works at a temperature ranging from 165K to 195K, the HgCdTe array of the IR channel has to work at a stable but adjustable temperature close to 80--85K during the acquisitions. The data produced by VIR are called hyperspectral images (or cubes) due to their structure: two spatial dimensions and one spectral dimension.\par
In the following Section \ref{VIR_data} we present the main characteristics of the VIR visible data set during the journey of Dawn around Ceres. Section \ref{INSTRU_EFFECTS} describes the VIS and the IR sensor temperatures behavior during the VIR acquisitions and how they affect the data. In Section \ref{sec_correction}, we present the VIR Arcturus observation and we address the question of the strategy adopted to define the correction as well as its application on the data. In Section \ref{RESULTS}, results of the correction are reported and the conclusions are presented in Section \ref{CONCLUSION}. 
\section{The VIR data set at Ceres}\label{VIR_data}
\begin{table*}
\caption{\label{Table1}Mission phases of Dawn at Ceres, chronologically sorted and during which VIR visible data were acquired. The fourth column presents the number of cubes processed and available in each phase. Differences are due to the occurrence of sky observation or corrupted data. The approximate minimum and maximum across-track resolutions are provided in the fifth column. The range of temperatures experienced by the visible and the infrared channel are reported in columns 6 and 7. The last column reports the correction factor used to correct the data set (see section \ref{App_CF}). Asterisks in the first column indicate which phases have been used to compute a correction factor.}
\resizebox{\textwidth}{!}{
\begin{tabularx}{17cm}{X X X X X X p{2.5cm} X}
\hline
\begin{tabular}{@{}l@{}}Mission\\ Phase\end{tabular} & \begin{tabular}{@{}l@{}}Start date\\yy-mm-dd\end{tabular} & \begin{tabular}{@{}l@{}}Stop date\\yy-mm-dd\end{tabular} & \begin{tabular}{@{}l@{}}Cubes\\(used/total)\end{tabular} & \begin{tabular}{@{}l@{}}Resolution\\ (m/pix)\end{tabular} & \begin{tabular}{@{}l@{}}VIS T (K)\end{tabular} & \begin{tabular}{@{}l@{}}IR T (K)\end{tabular} & \begin{tabular}{@{}l@{}}Correction\\factor\end{tabular} \\[5pt]
\hline \hline
CSA & 2015-01-13 & 2015-04-15 & 32/35 & 5000--10000+ & 170--189 & 80 & CSR \\[2.5pt]
CSR* & 2015-04-25 & 2015-05-07 & 75/75 & 3400--3500 & 169--186 & 80 & CSR \\[2.5pt]
CTS & 2015-05-16 & 2015-05-22 & 12/12 & 1300--1800 & 170--185 & 80 & CSR \\[2.5pt]
CSS* & 2015-06-05 & 2015-06-27 & 230/280 & 1000--1100 & 168--195 & 80 - 174/177 & CSS \\[2.5pt]
CSH* & 2015-08-18 & 2015-10-21 & 187/196 & 360--400 & 168--184 & 85 & CSH \\[2.5pt]
CSL* & 2015-12-16 & 2015-04-15 & 806/815 & 90--100 & 171--192 & 175 & CSL \\[2.5pt]
CXL & 2016-08-13 & 2016-08-27 & 136/139 & 90--100 & 169--189 & (160--172) - 175 & CSL \\[2.5pt]
CXG & 2017-01-27 & 2017-02-11 & 58/58 & 1900--2000 & 178--190 & 175 & CSL \\[2.5pt]
CXO & 2017-04-29 & 2017-04-29 & 18/19 & 4700--5100 & 170--186 & 90--175 & - \\[2.5pt]
C2E & 2018-06-09 & 2018-09-29 & 1/1 & 15--230 & 184--194 & 175 & CSL \\ \hline
\end{tabularx}
}
\end{table*}
The observations of Vesta and Ceres are grouped in different phases depending on the altitude of the spacecraft and/or of the scientific goals. Table \ref{Table1} describes the timeline and some characteristics of the ten mission phases during the Dawn mission at Ceres and for which visible data were acquired. Here we briefly describe the different mission phases in a chronological order. All the resolutions provided corresponds to across-track resolutions, which can differ to the along-track resolutions because of the exposure time and the spacecraft movement.\par
The CSA (Ceres Science Approach) phase corresponds to the Ceres approach and offers a full coverage of the surface at low resolution. The CSR (Ceres Science RC3 - Rotational Characterization 3) phase, aiming at study the rotation of Ceres, provides a complete coverage of the surface at intermediate resolution. CTS (Ceres Transfer to Survey) is an orbital transition phase with higher resolution than CSR but a low spatial coverage. It is followed by the survey phase, CSS (Ceres Science Survey), which covers almost all the surface with a resolution around 1000 m/pixel. The CSH (Ceres Science High Altitude Mapping Orbit or HAMO) phase covers latitudes between 40$^{\circ}$N and 65$^{\circ}$S with a resolution close to 350--400 m/pixel. The observation scheme during this phase is very regular due to the Dawn orbit. Data from this phase do not cover all the surface but provide a good coverage for the mid-latitudes and the areas of interest (e.g. bright spots, Occator, Dantu and Haulani craters). During the nominal mission, the closest distances to Ceres for a durable period of time for the visible channel were reached during the CSL (Ceres Science Low Altitude Mapping Orbit or LAMO) phase. The altitude ranges from 360 to 420 km from the Ceres' surface, allowing very detailed observations with a resolution close to 100 m/pixel. With more than 800 cubes acquired during CSL, a very good surface coverage (from 90$^{\circ}$N to 80$^{\circ}$S latitude) was achieved. After CSL, the mission entered the first Extended mission phase, called CXL (Ceres Extended LAMO). This orbit had the same characteristics of the previous CSL mission phase. At the end of CXL, the Dawn spacecraft orbit has been modified, thus entering in the CXG (Ceres Extended GRaND Orbit) phase, with a resolution of 2000 m/pixel and mostly devoted to GRaND acquisitions. This was followed by the CXO (Ceres Opposition) phase, with a lower resolution, that allowed the observation of the opposition effect on the Ceres surface\cite{2018_Ciarniello, 2019_Ciarniello}. Finally, VIR acquired the last VIS spectra during C2E (Ceres X2 Elliptical), the second extended mission phase. In that final phase, data with a resolution ranging from 230 m/pixel to 15m/pixel were acquired.\par
The data acquired by VIR are converted through the calibration pipeline \cite{2011_De_Sanctis,2013_VIR_Cal_Tech} from raw digital numbers (DN) to physical units of radiance. Then, they are converted in radiance factor following:
\begin{equation}
\nicefrac{I}{F} = \frac{\pi \times d^2 \times I}{F}
\end{equation}\par
Where $I$ is the calibrated radiance in units of $W.m^{-2}.\mu m^{-1}.sr^{-1}$, $F$ is the solar flux at 1 astronomical unit and $d$ is the heliocentric distance of Ceres at the time of the acquisitions.\par
The data used in this study are available on the PDS archive and corresponds to the LEVEL 1B version of the VIR data (\url{https://sbn.psi.edu/pds/resource/dawn/dwncvirL1.html}). A multiplicative factor related to the ground observation of Ceres and Vesta as well as a correction matrix is applied to all the data used here in order to correct the original deviation of the spectra and to reduce the spectral artifacts \cite{2016_Carrozzo}. For the purpose of this investigation, the data are also photometrically corrected \cite{2017_Ciarniello} to minimize the effects caused by the variations of the observation geometry.

\section{Instrumental effects on the VIR VIS data}\label{INSTRU_EFFECTS}
The visible data acquired at Ceres were used to study the characteristics of the surface using the map of the albedo distribution, the spectral slopes and other spectral parameters. However, in the data acquired during the different mission phases unexpected variations in spectral slopes and overall spectral shape occur independently of the scene observed, or the observational conditions in terms of geometrical factors, illumination conditions or temperature of the surface.\par
This behavior could have an instrument cause. Here we investigate the reasons of the unexpected changes in the VIS spectra in order to correct the data set.

\subsection{The visible detector temperature dependency on the visible data}\label{subsec_CCD_dependency}

\begin{figure}
\includegraphics{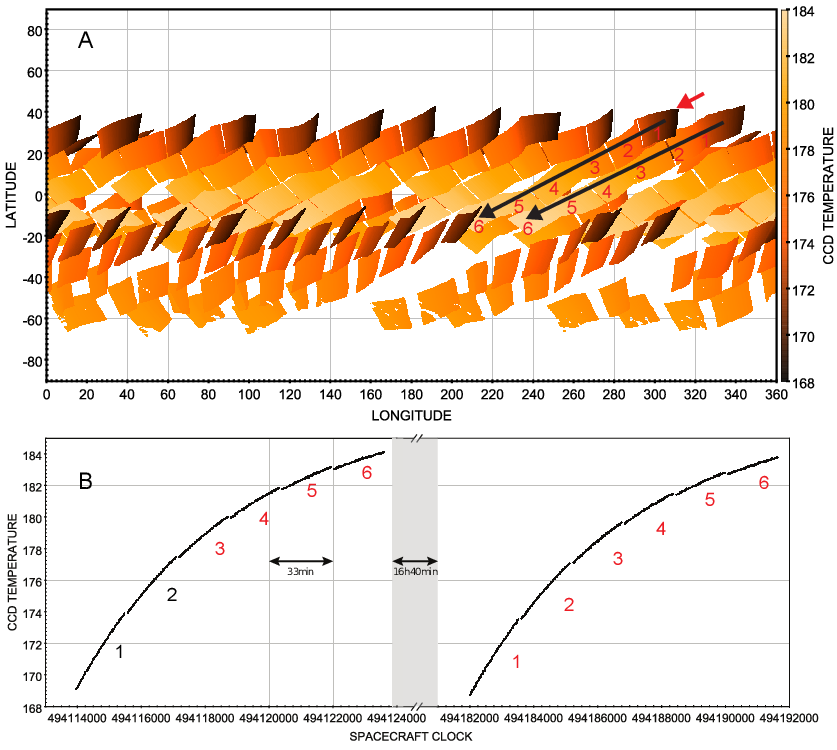}
\caption{\label{01_CCD_MAP} Panel A: temperature map of the CSH mission phase. Two sequences of acquisitions are marked by black arrows, with the beginning of the second indicated by a red arrow. Numbers indicate the acquisition order. Panel B: Evolution of the temperature for the two sequences pointed in Panel A. The total duration of each sequence is around 2 hours and 30 minutes. The VIS temperature is increasing from the beginning to the end of each sequence from 169K to 184K. The second sequence, indicated by the red arrow on the panel A, starts almost 17 hours after the end of the first one. During this time, the CCD sensor is able to cool down and temperature drops.}
\end{figure}

During a mission phase, the VIR observations are organized in sequences composed of hyperspectral cubes acquired in succession, without interruption. The acquisition schemes depend on the mission phase, thus mainly on the spacecraft orbit. Normally, during a sequence of acquisitions, the CCD detector of the VIS channel operates for hours consecutively. This effort leads to a temperature increase, while the cooling down is only possible when the detector is not operative, i.e. between two sequences. The range of temperatures experienced by the VIS sensor varies from 168K to 195K in the overall mission phases but may experience a smaller range within a single phase (see Table \ref{Table1}). Panel A of the Figure \ref{01_CCD_MAP} shows the map of the VIS sensor temperatures (VIS temperature hereafter) of the CSH phase providing an example of the typical effects experienced by the detector during a series of observation sequences. Along the first sequence of six cubes, numbered from 1 to 6 in black in Figure \ref{01_CCD_MAP}, the VIS temperature increases monotonically (also represented in the panel B of the Figure \ref{01_CCD_MAP}). When a new sequence starts after several hours (red arrow and numbering, panel A Figure \ref{01_CCD_MAP}), the VIS sensor has cooled down and the temperature increases until the last cube of the sequence.\par
\begin{figure}
\includegraphics{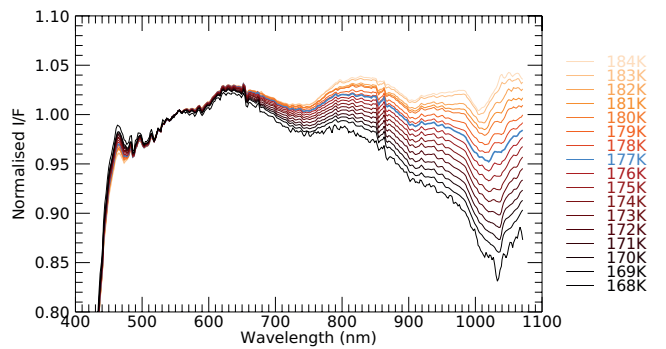}
\caption{\label{02_MED_SPEC} Median spectra of the CSH mission phase for the different VIS temperatures. The spectra are normalized at 550 nm. The reference spectrum (see Section \ref{CF_def}) is indicated in blue.}
\end{figure}

It has been observed that the changes of the VIS temperature has a strong impact on the global spectral slope. Figure \ref{02_MED_SPEC} represents the normalized median spectra of the CSH phase at different VIS temperatures. The spectral slope is regularly evolving from strongly negative to less negative or flatter while the VIS temperature is ranging from 168K to 184K respectively. In the CSH case, the variation of the radiance factor at 950 nm (after normalization at 550 nm) reaches $11\%$ between 168K and 184K. It represents $0.68\%/K$, which is close to the average value ($0.67\%/K\pm0.04$) observed for the most representative mission phases of the VIR VIS data set (CSL, CSR, CXL and CSH). Such spectral variations -- presented here in CSH but also observed during the other mission phases -- cannot be linked to the spatial distributions of these observations or to the diversity of composition across the Ceres' surface \cite{2016_Nathues,2016_Ammannito,2017_Ciarniello,2018_Carrozzo,2019_Frigeri}.
\subsection{The infrared channel temperature effect on the visible data}\label{subsec_IR dependency}
The VIS and IR channels composing the VIR instrument are decoupled and may observe independently, depending on the science objectives. During its operation, the IR channel is cooled down actively by a crycooler at a stable temperature close to 80--85K \cite{2011_De_Sanctis}. On the contrary, the VIS channel is cooled down passively thanks to a radiator and operates at higher temperatures \cite{2011_De_Sanctis}, typically from 165K to 195K as indicated in Table \ref{Table1}. When the IR channel is not in use, the cryocooler is switched off in order to preserve it. In that case, the IR temperature, as measured by the platinum resistors placed on the back side of the detector \cite{2011_De_Sanctis}, increases to reach progressively an equilibrium temperature around 175K (see also Table \ref{Table1}).\par
\begin{figure}
\includegraphics{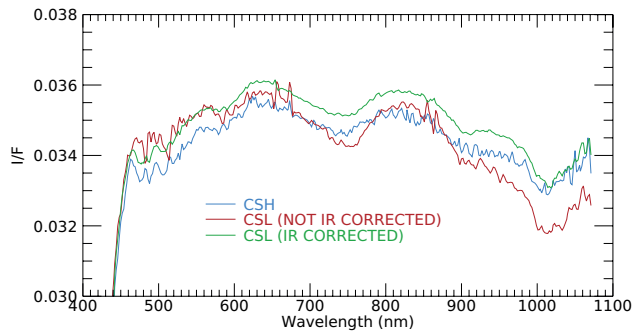}
\caption{\label{03_SHAPE_IR} Median of CSH (blue) and CSL (red and green) spectra (216 and 3414 respectively) acquired at the same location and under the same conditions of observation geometry and VIS temperature. The red CSL spectrum, which is not corrected for the high IR temperature effects, exhibits the distortion induced by this instrumental issue. The green spectrum is corrected for the IR temperature effects and is discussed in Section \ref{RESULTS}.}
\end{figure}

 
As illustrated by Figure \ref{03_SHAPE_IR}, we noticed different spectral behaviors for different mission phases (CSH and CSL, blue and red spectra of Figure \ref{03_SHAPE_IR}). The difference between the two spectral shapes cannot be explained by local variability across the surface or by different observation geometries, given that observations sampling the same region were selected under the same conditions. Looking at different spatial scales (global comparison between the mission phases or between different cubes), we tried to identify possibly relevant parameter(s) affecting the spectral response among different ones such as: the exposure time, the DN level, the spatial resolution, the local solar time and the instrumental temperatures measured during the acquisitions (telescope, radiator, spectrometer and IR detector temperatures). It seems that only the IR temperature is linked with the unusual spectral behavior observed.\par
In Figure \ref{03_SHAPE_IR}, we report the two median spectra of the same area (longitudes from 288.5$^{\circ}$E to 291$^{\circ}$E and latitudes from 3.9$^{\circ}$S to 0.4$^{\circ}$N). The blue spectrum is a median of 216 spectra coming from the cube 494868787 acquired during the CSH mission phase, while the red spectrum is a median of 3414 spectra of the cube 521987329 acquired during the CSL mission phase. As reported in Table \ref{Table1}, the IR temperature during CSH is 85K and 175K during CSL. In order to minimize photometric effects induced by observation geometry and the spectral ones, related to the above mentioned VIS temperature, we selected data acquired at the same VIS temperature (between 182K and 183.5K) and at the same phase angles (between 38$^{\circ}$ and 41.5$^{\circ}$). In addition, to further reduce the impact of these effects, we used photometrically corrected data \cite{2017_Ciarniello} and we applied the correction for the VIS temperature effect (see Section \ref{sec_correction} below).\par
The main differences among the blue and the red spectra of Figure \ref{03_SHAPE_IR} is the spectral slope which gets globally more negative as the IR temperature increases (CSL case, red spectrum). The shape is also modified: the CSL spectrum seems to be distorted, having a higher amplitude, especially in the range around 700--900 nm. While this spectral interval corresponds to a lower responsivity of the sensor, to date, it is not possible to establish whether it affects the resulting shape. According to the two CSL spectra of Figure \ref{03_SHAPE_IR}, the effect of the high IR channel temperature on the radiance factor measured at 950 nm is of the order of $4\%$.\par
With an equivalent number of spectra, the median spectra of CSH is less noisy than the one of CSL. We may suppose that the higher IR temperature, the more noise affects the VIS spectra. In addition, we also note an inversion of the spikes visible in the two spectra (e.g. between 460 nm and 530 nm and at 585 nm, 655 nm, 675 nm and 850--860 nm for the main ones). The later effect is indirectly due to the denoising procedure\cite{2016_Carrozzo} which corrects the data following an average trend. The spectra, once corrected, can be different in such structures depending on their initial shape. Finally, based on the comparison of Figure \ref{03_SHAPE_IR}, we exclude a systematic and major effect on the absolute albedo.\par
Having checked the influence of the parameters described above, we think that the main cause of the spectral distortion in the VIR VIS spectra is the thermal input coming from the IR sensor on the visible channel. Under this assumption, in the next section, we define an empirical correction that is designed to minimize the effects of the VIS and IR temperatures on the VIS channel, by adjusting the visible spectra to a common reference.
\section{Correction of the data}\label{sec_correction}

We showed that the spectral modifications observed in the VIR VIS data came from a) the VIS temperature variations during a sequence of acquisitions (Figure \ref{01_CCD_MAP} and \ref{02_MED_SPEC}) and b) the IR channel temperature (Figure \ref{03_SHAPE_IR}), which varies whether the IR channel is used or not. We illustrated that an increase of the VIS temperature implies a more negative spectral slope while a high IR temperature increases the amplitude of the spectral features and makes a slightly more negative spectral slope. Here we address the correction of the data, first by defining reference VIS and IR temperatures, then by calculating a correction factor, and finally by describing how to apply it.

\subsection{Absolute calibration and reference for the correction processes}\label{subsec_Absolute}

\begin{figure}
\includegraphics{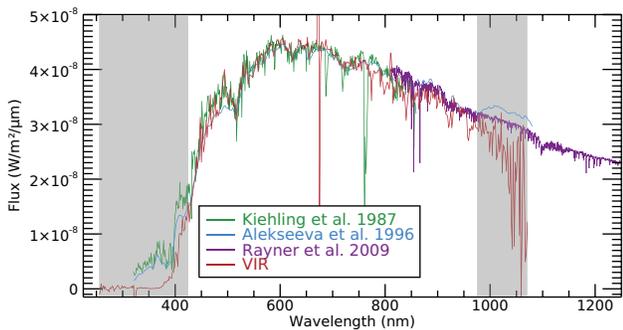}
\caption{\label{04_Arcturus}Comparison between Arcturus' spectra from VIR and different ground-based measurements from \citep{1987_Kiehling,1996_Alekseeva,2009_Rayner}. The spike around 675 nm in the VIR spectrum is due to an order sorting filter. Below 425 nm and after 975 nm, the signal in the VIR observation is below 60 DN and consequently too low to be reliable.}
\end{figure}

While several in-flight observations of stars have been performed during the Dawn mission, only one observation of Arcturus is suitable for the VIS channel calibration, being acquired at full spectral resolution. Figure \ref{04_Arcturus} reports this observation compared to three Arcturus' spectra acquired from Earth \cite{1987_Kiehling,1996_Alekseeva,2009_Rayner}. These ground-based observations have been chosen because their spectral resolutions are close to the one of VIR, therefore facilitating the comparison. In the ranges below 425 nm and above 975 nm (grey areas in Figure \ref{04_Arcturus}), the raw signal of the VIR observation is too low (below 60 DN) to be evaluated correctly. Consequently, those ranges are not considered here. In terms of absolute level, the agreement between the different ground-based observations and the VIR spectrum is very good. We only note a slight disagreement in the range 760--870 nm which could be explained by the presence of a minimum in the CCD responsivity in that range (see sec. \ref{subsec_IR dependency}). However, as previously mentioned, the impact of this minimum in the CCD responsivity on the spectral shape in that range is not confirmed to date.
The good match provided through this comparison validates the instrumental absolute calibration. This agreement implies that the VIS and IR temperatures of this observation can be used as a reference for the correction processes detailed in the next section. We then consider the two following assumptions:
\begin{enumerate}[noitemsep]
\item The VIS temperature -- of 177K during this acquisition -- will be considered as a reference for calculating the correction factor.
\item The IR temperature -- around 80K at that time and indicating that the cryocooler is in operation -- will be also considered as a reference.
\end{enumerate}
\subsection{Definition of the correction factors}\label{CF_def}

Due to the complexity of the VIR data set in terms of instrumental effects and observation parameters as well as for computational reason (see section \ref{App_CF}), it was not possible to derive a unique correction factor (CF) to be applied to the entire data set. At the same time, it was not possible to compute one correction for each mission phase, mainly due to lack of data for some of them. Given this, a minimum number of correction factors has been derived and applied to the different mission phases.\par
\begin{figure}
\includegraphics{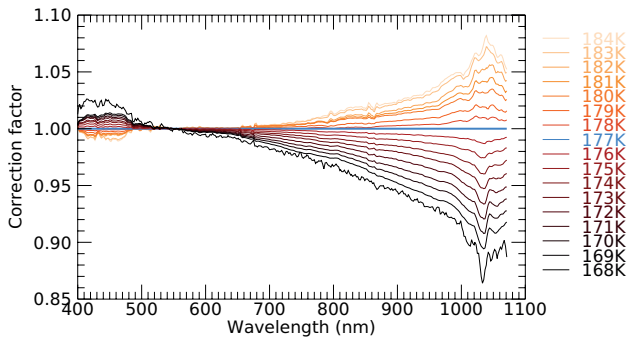}
\caption{\label{05_CF} Correction factor corresponding to the CSH mission phase. Its computation is based on the spectra of the Figure \ref{02_MED_SPEC} and Equation \ref{eqn:CF}. These data are normalized at 550 nm and 177K which is the temperature of reference.}
\end{figure}

As shown in Figure \ref{02_MED_SPEC}, we have grouped the observations by bin of 1K (VIS temperature) to build the correction factor. This correction factor, illustrated for the CSH mission phase in Figure \ref{05_CF}, is defined as the ratio between the normalized median spectra at a given bin of VIS temperature and the normalized median spectrum as derived from the CSH data set, at the VIS temperature of 177K and at the IR temperature of 85K, i.e.: 
\begin{equation}
\label{eqn:CF}
CF_{\lambda,T_{VIS},T_{IR}}=\frac{(\nicefrac{I}{F})_{\lambda,T_{VIS},T_{IR}}}{(\nicefrac{I}{F})_{CSH,\lambda,T_{VIS}=177K,T_{IR}=85K}}
\end{equation}\par
Where $I/F$ at the numerator is the median radiance factor normalized at 550 nm and for a given bin of VIS temperature of the mission phases for which a correction factor is computed; $(I/F)_{CSH}$ at the denominator is the median radiance factor normalized at 550 nm at a VIS temperature of 177K and an IR temperature of 85K of the CSH mission phase; $\lambda$ is the wavelength; $T_{VIS}$ is the VIS temperature and $T_{IR}$ is the IR temperature.\par
The CSH spectrum acquired at 177K ($T_{VIS}$) and 85K ($T_{IR}$), colored in blue in the Figure \ref{02_MED_SPEC}, is considered as the reference spectrum because the 285319 spectra used to build it are distributed across the whole range of longitudes and at mid-latitudes (between 60$^{\circ}$S to 30$^{\circ}$N). This insures a good representation of Ceres' surface composition, avoiding both bias from uneven sampling and extreme observation geometries. Also, even if the IR temperature of reference is 80K while the one of the CSH is 85K, we note that no differences are observed between the median spectra of the whole data of CSH (IR temperatures around 85K) and CSR (IR temperatures around 80K), which makes possible the choice of CSH as a reference despite the slightly different IR temperature.\par
The production of the correction factor relies on the following hypotheses: 
\begin{enumerate}[noitemsep]
\item Since the spectra grouped in bins of VIS temperature come from different locations of the surface with possibly different albedos due to variations of composition or residuals of the photometric correction, they have been normalized at 550 nm. The implication of this first hypothesis is that the radiance factor at 550 nm will have the same value before and after the correction, independently from the VIS temperature. It implies also that around this wavelength, the correction will be lower than potentially needed and prevent us to correct for effects on the radiance factor absolute level. 
\item A normalization relative to the VIS temperature is also needed since we have to assume that at a given temperature, the spectrum does not need correction. As described in the section \ref{subsec_Absolute}, the temperature of 177K has been chosen in light of the absolute calibration tested with Arcturus' observation.
\item In the same way, we assume that during the VIR regular operating mode, the cryocooler is switched on, which implies that the IR temperature is close to 80K-85K. This assumption appears to be solid thanks to the quality of the VIR Arcturus observation, performed when also the IR was operating, in comparison to the ground based ones.
\end{enumerate}
Following Equation \ref{eqn:CF}, the correction factor for a single mission phase (e.g. CSH, CSS…) describes a set of vectors, one for each VIS temperature bins, covering the VIR VIS spectral range. Figure \ref{05_CF} represents the CSH correction factor based on the spectra of Figure \ref{02_MED_SPEC}.\par
In Section \ref{subsec_IR dependency} we described the effect of an IR temperature around 175K and we explained the choice of the reference at 80--85K. While the VIS temperature continuously evolves during the acquisition sequences, the IR temperature does not change and is constant during each mission phase. The only exceptions are CXO and a part of CSS, for which the IR channel was operating at the beginning of the sequences and then switched off at the end, as described later. By definition, the use of the reference spectrum of CSH within the correction factor allows us to correct the distortion induced by high IR temperature when present (e.g. CSL) while it will have no effect for other mission phases acquired with an IR temperature of 80--85K (e.g. CSR).\par
Finally, this correction factor is adequate to correct both the VIS temperature effect and the IR temperature distortion while conserving the relative variability of each individual spectrum. It allows the comparison and mapping of different cubes acquired under variable instrumental conditions and provides a relatively homogeneous result.
\begin{figure*}
\includegraphics{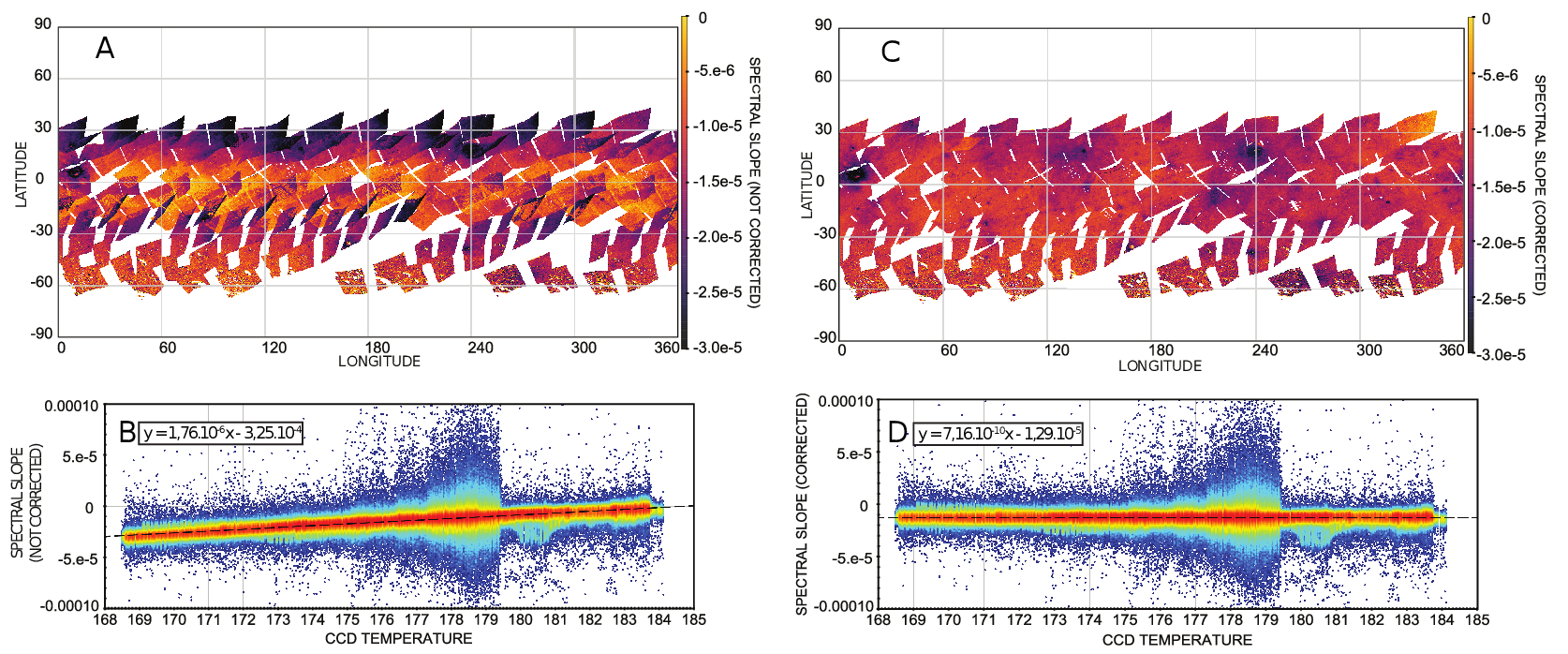}
\caption{\label{06_SLOPE_BEFORE_AFTER} Panel A: map of the spectral slope (630--950 nm) for the CSH mission phase without the correction of the VIS temperature effect. Panel B: distribution of the spectral slope (630--950 nm) as a function of the VIS temperature for the data of Panel A. A clear trend is visible in both cases due to the increase of the VIS temperature (see also Figure \ref{01_CCD_MAP}). The interpretation of the map is not possible without a correction of the data. Panel C: map of the spectral slope after the correction of the VIS temperature effect. Panel D: distribution of the spectral slope in function of the VIS temperature after the correction. Once the correction is applied, no trend can be noted in the distribution of panel D. Intrinsic spectral variability of Ceres’ surface relevant morphological features can be clearly recognized in the map of the panel C, e.g.: Haulani close to the longitude 0--20$^{\circ}$E, latitude 0--20$^{\circ}$N; Dantu around the longitude 140$^{\circ}$E, latitude 20$^{\circ}$N; Occator on longitude 240$^{\circ}$E, latitude 20$^{\circ}$N.}
\end{figure*}
\subsection{Application of the correction factors\label{App_CF}}

The correction factor to be applied for each spectrum must take into account the VIS temperature measured at the time of the acquisition. Those VIS temperatures, as well as the IR temperature, are provided by the housekeeping parameters of VIR. As an example, for a given pixel, the VIS temperature can be 172.34k, while the correction factor is available for 172K and 173K. We then calculate an interpolation -- made with the IDL routine INTERPOLATE -- which allows us to determine the value of the correction factor for this effective VIS temperature. Then, the correction of the data is performed by dividing each spectrum by the correction factor interpolated at the corresponding VIS temperature.\par
As indicated in Table \ref{Table1}, different correction factors have been computed for the CSR, CSS, CSH and CSL mission phases. Given the VIS temperature ranges and CSS IR temperatures distribution (see below), this is the minimal number of correction factors that can be used. To correct the CSA, CTS, CXL, CXG and C2E mission phases, the correction factor of other mission phases has been used. The choice is based on the VIS temperature range spanned in the different sequences and the corresponding IR temperature. In order to apply the correction factor properly, the VIS temperature range from which it is derived should encompass the one of the mission phase that has to be corrected. Simultaneously, the IR temperature of the data to be corrected must match the data used to compute the correction factor. For example, the CSL correction factor will be used to correct the CXL, CXG and C2E data, for which IR temperature of 175K corresponds to a spectral shape similar to the red spectrum of the Figure \ref{03_SHAPE_IR}. On the other hand, we cannot use the CTS or the others mission phases, which have IR temperature of 80K and a spectral shape similar to the blue spectrum of the Figure \ref{03_SHAPE_IR}. The Table \ref{Table1} indicates also which correction factor must used to correct a given mission phase.\par 
Some exceptions or particular cases are nonetheless worthy to be detailed below, because if they are ignored, they can lead to imprecise correction causing artifacts on final maps.
\begin{enumerate}[(a)]
\item The CXL mission phase is corrected with the CSL correction factor, because both have a high IR temperature. However, the CXL VIS temperature range is larger than the one of CSL since the CXL's starts at 169K while the CSL's at 171K (see Table \ref{Table1}). The CXL data with a VIS temperatures comprised between 169K and 171K are then corrected by the CSL correction factor at the VIS temperature of 171K, without extrapolation. Because it impacts only $0.2\%$ of the VIS CXL data set, this can be neglected. We also note that a fraction of the CXL data are acquired while the IR temperature is comprised between 160--172K. Therefore, the CSL correction may not be fully adequate in that case and some residuals may be observed even after the application of the correction. While this subset of data represents $18\%$ of the CXL mission phase, it represents only $1.3\%$ of the whole VIR VIS data set and then can be discarded for mapping purposes, if necessary.
\item The CSS data set represents a peculiar case. One part of the corresponding data set has been acquired with an IR temperature around 80K (IR channel/cryocooler switched on), and the other part has been acquired around 175K (IR channel/cryocooler switched off). However, for $\sim92\%$ of the observations acquired at an IR temperature of 80K the VIS temperature was below 182K (which represents $\sim66\%$ of CSS data), while the rest of the data, acquired at an IR temperature of 175K, the VIS temperature were higher than 182K. Consequently, a part of the CSS data shows a CSH-like spectrum and the other part exhibits a CSL-like spectrum (blue and red spectrum of Figure \ref{03_SHAPE_IR} respectively). Following the definition of the correction factor, the spectrum of reference, acquired at a VIS temperature of 177K, is not affected by the high IR temperature effect and the CSS correction factor is then behaving well for the two parts of its data set. However, the particularity of the CSS mission phase implies that its correction factor cannot be used to correct the others missions phases.
\item For the CXO mission phase, which represents only $0.6\%$ of the whole VIR VIS data set, the IR temperature increased progressively from 90K to 175K during all the acquisitions because the IR cryocooler has been turned off just before the VIS channel started to operate. This configuration could have provided a chance to characterize the variation of the spectral response as a function of the IR temperature. However, this mission phase has been dedicated to the study of the opposition effect, and during the VIS sequence, while the IR temperature was raising, the phase angle of the observations progressively increased from $\sim 0^{\circ}$ to $\sim 6.6^{\circ}$. This introduced an unwanted correlation between the phase angle and the IR temperature. Phase angle variations may in principle introduce spectral variability, thus, in this case, we are not able to disentangle intrinsic spectrophotometric effects from the spurious ones introduced by the IR temperature variability (which is of the order of $4\%$, see Section \ref{RESULTS}). Given this, we choose to not apply or fix any correction factor for this mission phase.
\end{enumerate}
\begin{figure}
\includegraphics{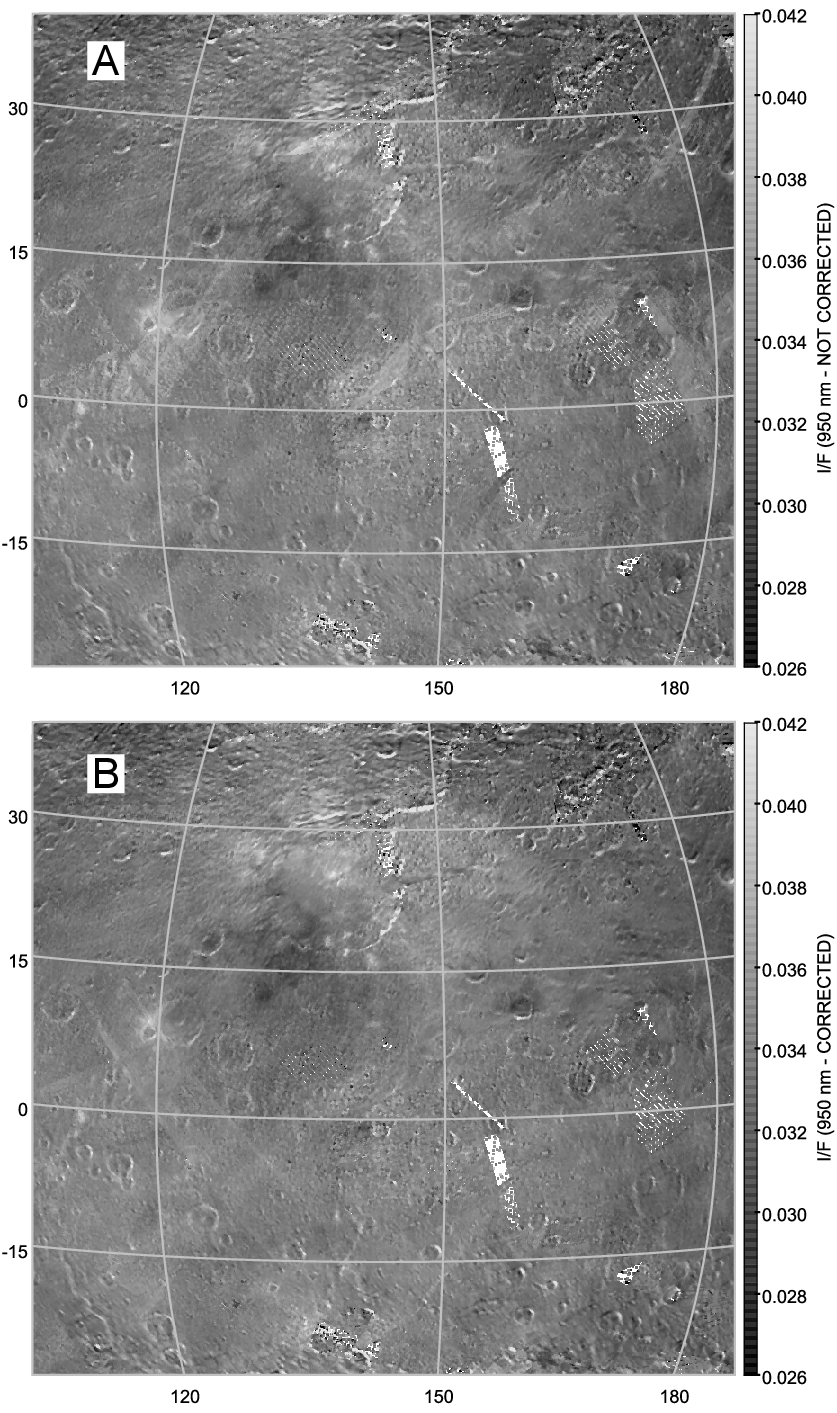}
\caption{\label{07_IoF_DANTU_BEFORE_AFTER}Orthographic projection of the Dantu region using the radiance factor at 950 nm, before the correction of the VIS and IR temperature effects (panel A) and after the correction (panel B). In panel A, the data are already photometrically corrected. The data used come from the CSH, CSS, CSR and CTS mission phases.}
\end{figure}
\begin{figure}
\includegraphics{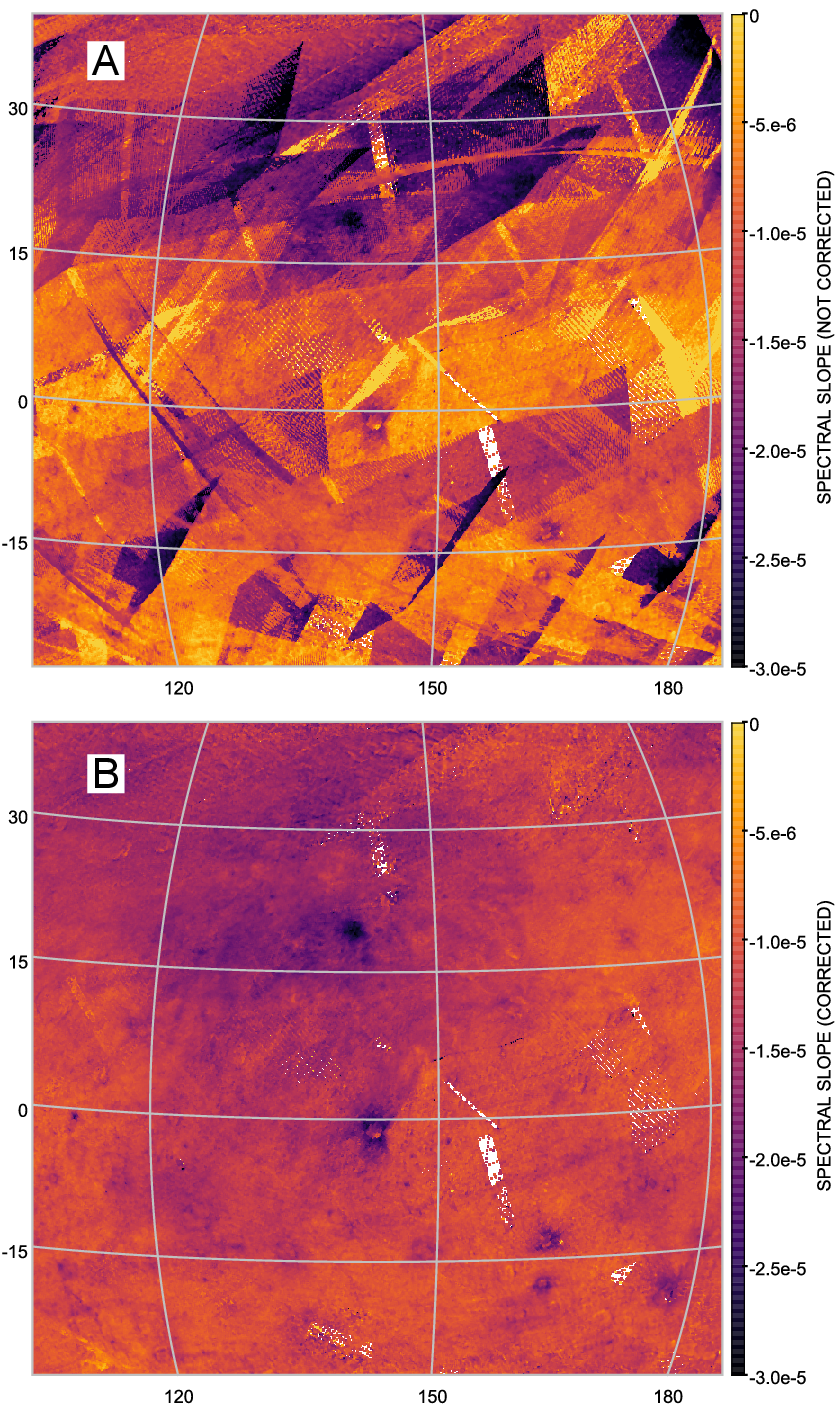}
\caption{\label{08_SLOPE_DANTU_BEFORE_AFTER}Orthographic projection of the Dantu region using the spectral slope (see the text for the definition), before the correction of the VIS and IR temperature effects (panel A) and after the correction (panel B). In panel A, the data are already photometrically corrected. The data used come from the CSH, CSS, CSR and CTS mission phases.}
\end{figure}

\section{Results}\label{RESULTS}

Because the changing VIS temperature affects mostly the spectral slope, we define the following slope parameter in order to illustrate the effectiveness of the correction: 
\begin{eqnarray}
\label{eqn:SLOPE}
	S_{max(620|650)\text{--}950nm}=\nonumber\\
    \frac{(\nicefrac{I}{F})_{950nm}-(\nicefrac{I}{F})_{max(620|650)nm}}{(\nicefrac{I}{F})_{max(620|650)}\times(9500{\text{\normalfont\AA}}-max(6200|6500){\text{\normalfont\AA}})}
\end{eqnarray}
Where $S_{max(620|650)\text{--}950}$ ($S$ hereafter) is the spectral slope, computed between the radiance factor at 950 nm $((I/F)_{950nm})$ and the position at which the maximum occurs in the interval 620--650 nm $((I/F)_{max(620|650nm)})$.
Figure \ref{06_SLOPE_BEFORE_AFTER} presents the maps and the distributions of the spectral slope as defined in Equation \ref{eqn:SLOPE} before (panels A and B) and after (panels C and D) the application of the correction factor for the CSH mission phase. The maps presented were obtained by projecting the footprint of each single pixel on an equidistant cylindrical projection. All overlapping footprints are merged by calculating the median.\par
After correction, for $98.2\%$ of the 3 219 440 CSH processed spectra the spectral slope is distributed between $S=-3.10^{-5}  k{\text{\normalfont\AA}}^{-1}$ and $S=0  k{\text{\normalfont\AA}}^{-1}$ (with a median at $S=1.26\times10^{-5}  k{\text{\normalfont\AA}}^{-1}$). This range is chosen for the color scale of the Figure \ref{06_SLOPE_BEFORE_AFTER} even if some locations such as Haulani (longitude $\sim$10$^\circ$E - latitude $\sim$10$^\circ$N), and Occator crater (longitude $\sim$240$^\circ$E - latitude $\sim$20$^\circ$N) or some pixels at high incidence angles and at southern latitudes show values exceeding these limits. Panels B and D of Figure \ref{06_SLOPE_BEFORE_AFTER} show a linear fit (least square) across the distribution of the spectral slope against the VIS temperature. The fit is calculated in the restricted spectral slope range $-10^{-4}\text{--}10^{-4} k{\text{\normalfont\AA}}^{-1}$ in order to exclude a minority of outliers identified by their very low DN level and extreme slope values. After correction, the two excluded subsets represent only 1684 pixels over more than 3 million for the fit of the panels B and D in Figure \ref{06_SLOPE_BEFORE_AFTER}. The computed slope of the distribution is reduced from $-1.76\times10^{-4}\%  k{\text{\normalfont\AA}}^{-1}.\text{K}^{-1}$ to $-7.16\times10^{-10}\%  k{\text{\normalfont\AA}}^{-1}.\text{K}^{-1}$, indicating that the correction works well. The final value of the spectral slope is $S=-1.29\times10^{-5} k{\text{\normalfont\AA}}^{-1}$, which is very close to the median of the distribution.\par
The correction makes it possible to analyse specific regions or features, using different mission phases and resolutions. This is illustrated by the maps of Figures \ref{07_IoF_DANTU_BEFORE_AFTER} and \ref{08_SLOPE_DANTU_BEFORE_AFTER} centered in the South region of Dantu. The maps are orthographic projections of the data available in CSH, CSS, CSR and CTS mission phases. As opposed to Figure \ref{06_SLOPE_BEFORE_AFTER}, the pixels are represented as points and not as polygons of their surface footprints. However, the density of pixels as well as their spatial resolution in that region guarantees good coverage (except in few locations, e.g. $\sim160^\circ$E of longitude and $5^\circ$S of latitude) and avoid artifacts in the map due to this mapping technique \cite{2019_Rousseau_a}. As in Figure \ref{06_SLOPE_BEFORE_AFTER}, overlapping pixels are merged by calculating the median. The data used in the panel A of the Figures \ref{07_IoF_DANTU_BEFORE_AFTER} and \ref{08_SLOPE_DANTU_BEFORE_AFTER} are already photometrically corrected but not corrected for the temperature effect. In the temperature-corrected observations of the radiance factor at 950 nm, a small, although significant improvement can be noted with respect to the non-corrected ones (panel B in Figure \ref{07_IoF_DANTU_BEFORE_AFTER}). For the spectral slope, the improvement is more substantial (panel B in Figure \ref{08_SLOPE_DANTU_BEFORE_AFTER}). For example, the region around Dantu crater (longitude $\sim140^\circ$E, latitude $\sim20^\circ$N) was not distinguishable before the correction and it is clearly visible after. In addition, the ejecta ray coming from the Occator crater\cite{2019_Nathues} can be recognized in the map of the corrected spectral slope (panel B Figure \ref{08_SLOPE_DANTU_BEFORE_AFTER}) and it can be followed across the map on the whole range of longitudes and from latitude $0^\circ$N to about $15^\circ$S.\par
In Figure \ref{03_SHAPE_IR}, the CSL green spectrum is corrected with the correction factor based on the CSH reference. It allows the removal of the distortion caused by the high IR temperature and the retrieval of the right spectral slope and shape. A homogeneous behavior in the noise is also retrieved between CSH and CSL, which was not the case without the IR correction as explained in section \ref{subsec_IR dependency}. However, due to the normalisation in the correction process (see Section \ref{eqn:CF}), a systematic difference in albedo is observed between the CSH and CSL spectra of Figure \ref{03_SHAPE_IR} (blue and green respectively), which we do not expect here since the data has been corrected and, in particular, acquired on the same area. This systematic residual cannot be corrected -- by definition of the correction factor --  and points out the limit of our correction when the IR channel temperature is high. However, in the range 460--950 nm, this residual is constant and of the order of $\sim 1\%$ (by comparing the CSH spectrum and the CSL green spectrum). It is consequently a minor side effect compared with the original distortion induced by the high IR temperature which is around $\sim 4\%$ at 950 nm (by comparing the CSL red and green spectra before and after the IR correction).

\FloatBarrier

\section{Conclusion}\label{CONCLUSION}

Data acquired at Ceres by the visible channel of the VIR imaging spectrometer are affected by the VIS temperature variations during different sequences of acquisitions. In addition to the variation of the spectral slope induced by this first parameter, the IR temperature, linked to the use of the IR channel/cryocooler, plays a secondary role on the spectral slope but also on the global shape of the spectra. We developed an empirical correction and we applied it to the whole Ceres data set. The correction uses a unique spectral reference, chosen to be representative of the surface of Ceres and in the optimum ranges of the VIS and IR temperatures. The spectral reference has been defined according to the quality of the match existing between the VIR and ground-based observations of Arcturus. This consistency allowed to define the VIS and IR temperatures of reference and to highlight the accuracy of the absolute calibration. The correction applied to the VIR visible data of Ceres proved necessary to interpret and compare different set of observations and perform mapping. The spectral slope and shape are also more reliable because the spurious spectral variability has been minimized, both at global and local scale. In this respect, the final corrected data set of the VIR visible channel is now homogeneous within a single mission phase and also among different ones.\par
Data acquired at Vesta by the VIS channel of VIR are also affected by the VIS temperature variations. However, because the IR channel/cryocooler was always turned on during the VIS channel acquisitions (except for a very limited part of the data set), the Vesta data are not affected by the spectral distortion due to high IR channel temperature.\par

\section*{Supplementary Material}
See supplementary material for the figures of the median spectra and correction factors of the CSR, CSS and CSL mission phases.

\begin{acknowledgments}
VIR is funded by the Italian Space Agency-ASI and was developed under the leadership of INAF-Istituto di Astrofisica e Planetologia Spaziali, Rome-Italy (Dawn ASI INAF I/004/12/0). The instrument was built by Selex-Galileo, Florence-Italy. The authors acknowledge the support of the Dawn Science, Instrument, and Operations Teams. The author made use of TOPCAT (Tools for OPerations on Catalogues And Tables) \cite{2005_Taylor} for a part of the data analysis and figure production. We are grateful for the anonymous reviewers for their constructive suggestions that helped improve this manuscript.
\end{acknowledgments}
%
\bibliography{01_DRAFT_RSI.bbl}
\end{document}